\documentclass[twocolumn,showpacs,preprintnumbers,amsmath,amssymb]{revtex4}

\usepackage{graphicx}
\usepackage{dcolumn}
\usepackage{bm}

\begin{document}

\title{Thermally induced spin flips above an atom chip}

\author{M. P. A. Jones, C. J. Vale, D. Sahagun, B. V. Hall and E. A.
Hinds}
\email{ed.hinds@imperial.ac.uk}
\affiliation{Blackett Laboratory, Imperial College, London SW7 2BW, United Kingdom}%
\date{\today}

\begin{abstract}
We describe an experiment in which Bose-Einstein condensates and
cold atom clouds are held by a microscopic magnetic trap near a
room temperature metal wire 500\,$\mu$m in diameter. The ensemble
of atoms breaks into fragments when it is brought close to the
ceramic-coated aluminum surface of the wire, showing that
fragmentation is not peculiar to copper surfaces. The lifetime for
atoms to remain in the microtrap is measured over a range of
distances down to $27\,\mu$m from the surface of the metal. We
observe the loss of atoms from the microtrap due to spin flips.
These are induced by radio-frequency thermal fluctuations of the
magnetic field near the surface, as predicted but not previously
observed.
\end{abstract}

\maketitle

The ability to control cold atom clouds in microscopic magnetic
traps \cite{weinstein95,vuletic98,fortagh98} and waveguides
\cite{mueller99,dekker00,key00} has created the  new field of
miniaturized atom optics \cite{hindsreview99,folmanreview02}. With
the use of microstructured surfaces (atom chips) it becomes
possible to control cold atoms on the $\mu$m length scale and to
anticipate the construction of integrated atom interferometers
\cite{hinds01,haensel01,andersson02}. Ultimately there is the
possibility of controlling the quantum coherences within arrays of
individual atoms for use in quantum information processing
\cite{calarco00,horak02}. For these kinds of applications it is
important to avoid fluctuating or inhomogeneous perturbations,
which tend to destroy the quantum coherences.

Clouds within 100\,$\mu$m of a current-carrying wire and cooled
below a few $\mu$K have recently revealed three surface-related
decoherence effects. First, the clouds break into fragments along
the length of the wire as a result of a corrugated trapping
potential \cite{fortagh02,leanhardt02}. The corrugations are
caused by a small spatially alternating magnetic field {\it
parallel} to the wire \cite{kraft02}, which is presumably due to a
small transverse component of the current. The second effect is
heating of the cloud \cite{haensel01,fortagh02} due to
audiofrequency technical noise in the currents that form the
microtrap, which cause it to shake. Finally, trapped atoms are
lost \cite{haensel01,fortagh02} through spin flips induced by
radio frequency technical noise in the wire currents. Some of
these effects have recently been elucidated by Leanhardt {\it et
al.} through a comparison of magnetic and optical traps near a
surface \cite{leanhardt02a}.

In addition to these essentially technical decoherence effects,
there is a more fundamental limitation associated with the thermal
fluctuations of the magnetic field. When the cold atoms are far
from any surfaces in their room-temperature environment, they
interact with the blackbody radiation field. This has very little
power at the resonance frequencies of the atoms, making the
thermalization times exceedingly long and the decoherence effects
correspondingly small. However, atoms trapped some tens of $\mu$m
above a metal interact with the thermally fluctuating near field
of the surface, whose spectrum is very different from the
blackbody spectrum. Recent calculations \cite{henkel99} have shown
that spin flips induced by this near field can cause atoms to be
ejected from a magnetic trap in less than a second. In this letter
we present measurements of this fundamental effect, which, in the
presence of technical losses, has not previously been accessible
\cite{fortagh02,leanhardt02a}.

\begin{figure}
\includegraphics[width=3.3in]{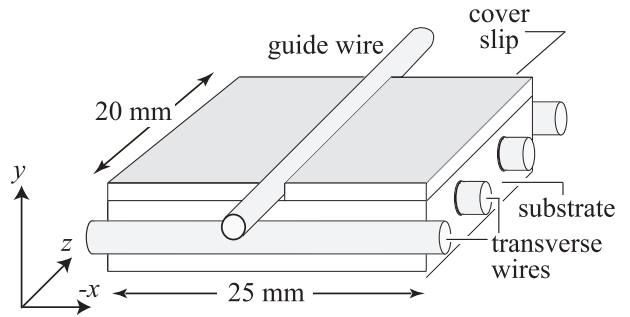}
\caption{Construction of the atom chip (not to scale). Atoms are
trapped near the surface of the guide wire.} \label{fig:chip}
\end{figure}

A diagram of the atom chip used to form our microtrap is shown in
figure \ref{fig:chip}.  The main wire is a 500\,$\mu$m diameter
guide wire along the {\it z}-direction. It consists of a central
core of copper with $185\,\mu$m radius, a $55\,\mu$m thick
aluminum layer and a $10\,\mu$m thick ceramic sheath. This wire is
glued by high-vacuum epoxy (Bylapox 7285) into the
$200\,\mu$m-deep channel formed by a glass substrate and two glass
cover slips. Below the guide wire there are four transverse wires,
$800\,\mu$m in diameter. The cover slips are coated with 60\,nm of
gold so that they reflect 780\,nm light. In order to make a
Bose-Einstein condensate in the microtrap we first collect
$^{87}$Rb atoms using a magneto-optical trap whose beams are
reflected from the gold surface. This MOT collects
$1\,\times\,10^8$ atoms at a height of 4\,mm above the surface and
cools them to $50\,\mu$K. The MOT is pulled down to a height of
1.3\,mm by passing a current of 3.2\,A through the guide wire and
adding a uniform magnetic field $B_x$ of 6\,G along the {\it
x}-direction.  This compresses the cloud into a cylindrical shape
and increases the phase space density of the atoms to $2 \times
10^{-6}$ \cite{jones02}.

The light and the anti-Helmholtz coils of the MOT are then
switched off and the atoms are optically pumped into the $|F,m
\rangle = |2,2 \rangle$ state. We collect $2\,\times\,10^7$ of
these atoms in the magnetic guide formed by the guide wire (8\,A
along {\it z}), and the transverse bias field $B_x$ (10\,G along
{\it x}). Axial confinement is provided by the inner transverse
wires (15\,A each along {\it -x} ), and the outer transverse wires
(15\,A each along {\it x} ). The field at the center of this trap
is partly cancelled by an axial bias field $B_z$ ($6\,$G along
{\it z}). Next, the trap is adiabatically compressed over 0.5\,s
by increasing $B_x$ and $B_z$ to $29\,$G and $11\,$G respectively,
and reducing the guide current to $6.9\,$A. This brings the trap
to a distance of 225\,$\mu$m from the wire and raises the radial
and axial trap frequencies to 840\,Hz and 26\,Hz. The elastic
collision rate is now $\sim 54$\,s$^{-1}$, which is high enough
for forced rf evaporative cooling to be efficient. We sweep the rf
frequency logarithmically over 12.5\,s from 13\,MHz to a final
frequency near 600\,kHz. This cools the cloud down to well below
the 380\,nK critical temperature for Bose-Einstein condensation
and produces up to $5 \times 10^4$ atoms in the condensate
\cite{jones02}.

We find that the number of atoms in the microtrap decays
exceedingly slowly with time. We cannot leave the trap on long
enough to measure the decay precisely because the vacuum
feedthroughs that carry the trap currents overheat, causing a
sudden increase of pressure after $\sim 20$\,s. However, the
lifetime is well in excess of 100\,s. For many of our measurements
we stop the evaporation after 6\,s at a temperature in the range
of $1-5\,\mu$K. These thermal clouds also have lifetimes well over
100\,s.

\begin{figure}
\centering
\includegraphics[width=3.4in]{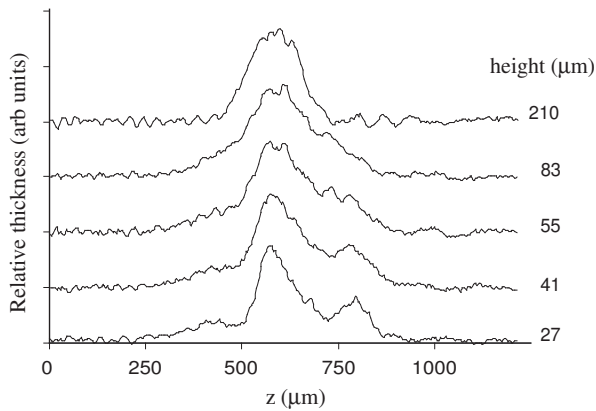}
\caption{Optical thickness of a $2\,\mu$K atom cloud measured by
absorption imaging.
 As the cloud approaches the
wire it begins to fragment along the {\it z}-direction. This
effect has previously been seen above copper wires.}
\label{fig:fragments}
\end{figure}

To bring the atoms closer to the surface, we smoothly reduce the
current flowing in the guide wire during the last $1\,$s of
evaporation. The arrival of the atoms at the desired height
coincides with the end of the evaporation ramp. The cloud is then
viewed on a ccd camera using absorption imaging with the probe
beam propagating along the {\it x}-direction. At distances above
$100\,\mu$m from the surface the axial profile of the cloud (i.e.
along the {\it z}-direction) is Gaussian, as shown in figure
\ref{fig:fragments}. However, as the cloud approaches the surface
of the wire, it breaks into fragments, similar to those observed
by other groups close to copper wire \cite{fortagh02,leanhardt02}.
At a height of $27\,\mu$m, the potential wells responsible for the
fragmentation shown in figure \ref{fig:fragments} are $\sim
1\,\mu$K deep and are due to a magnetic field parallel to the wire
\cite{kraft02} that alternates between $\sim\pm10\,$mG. Relative
to the expected field of the wire this anomalous part is $\sim 3
\times 10^{-4}$. The depth of these wells increases strongly as
the cloud approaches the aluminum surface, even though it is still
almost $100\,\mu$m away from the copper. This suggests that
fragmentation is not peculiar to copper but applies equally to
currents flowing in aluminum. Compared with references
\cite{fortagh02,leanhardt02}, the current density in our wire is
much lower, making it unlikely that the fragmentation is a result
of current instability at high density, as has been suggested
\cite{kraft02}. It remains an open question whether this
phenomenon is due to some fundamental physics, such as arrangement
of the spins in the conduction electrons \cite{fortagh02,kraft02},
or simply to imperfections in the geometry of the wire.

\begin{figure}
\centering
\includegraphics[width=3.4in]{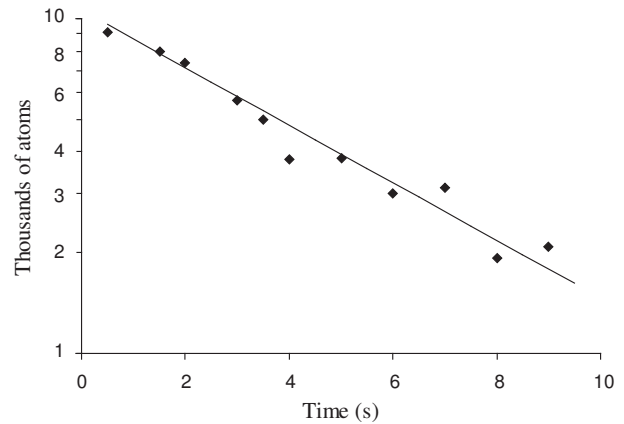}
\caption{Number of atoms remaining in the microtrap versus time at
a distance of $29\,\mu$m from the conducting  surface of the
wire.}
\label{fig:decaycurve}
\end{figure}

Figure \ref{fig:decaycurve} shows the total number of atoms in the
(fragmented) microtrap versus time, measured at a height of
29(4)\,$\mu$m above the aluminum surface of the wire. A small
number of atoms is used in order to avoid the unnecessary
complication of 3-body loss from the trap. This curve yields a
lifetime of 5.1(0.3)\,s - very much shorter than the lifetime far
from the wire. The increased loss rate is due to spin flip
transitions $|2,2\rangle \rightarrow |2,1\rangle \rightarrow |2,0
\rangle$ driven by thermal fluctuations of the magnetic field
close to the wire. The resonant frequency $f_0$ for this spin flip
transition is $\frac{1}{2}\mu_B B_0/h$, where $\mu_B$ is the Bohr
magneton, $h$ is Planck's constant, and $B_0$ is the magnetic
field at the center of the trap, where the cold atoms reside. This
field is produced by the combination of the transverse wires and
the axial bias field $B_z$, and it is controlled by adjusting
$B_z$.

Since the evaporative cooling uses these same spin flip
transitions, we are able to determine the transition frequency
$f_0$ by making several evaporations with different values of the
final rf frequency. The final temperature, as measured by the mean
square length of the thermal cloud, is linearly related to this
final frequency. Extrapolation to zero cloud length yields the
spin flip frequency $f_0$ of atoms at the center of the trap. The
measurements shown in figure \ref{fig:decaycurve} were taken with
$f_0=1.8\,$MHz, corresponding to a field of $B_0=2.6\,$G.

\begin{figure}
\centering
\includegraphics[width=3.3in]{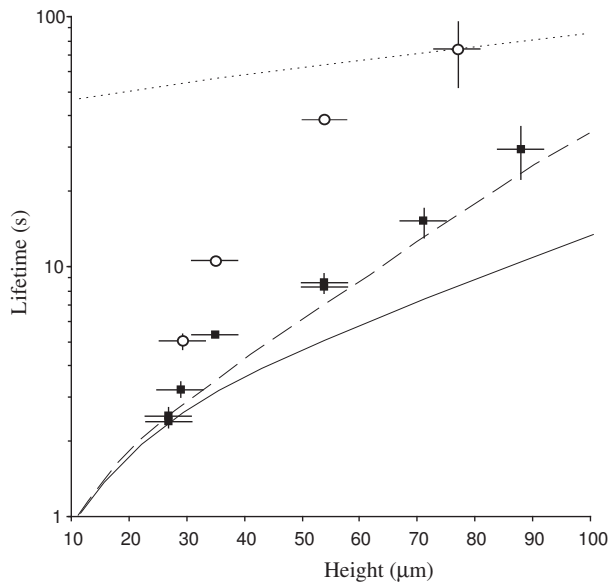}
\caption{Lifetime of cold trapped atoms versus distance from the
aluminum surface of the wire. Squares (circles): spin flip
frequency $f_0=560\,$kHz ($f_0=1.8\,$MHz). Dotted line: expected
scaling for technical noise. Solid (dashed) line: theoretical
prediction for lifetime above a thick slab of aluminum with
$f_0=560\,$kHz ($f_0=1.8\,$MHz)} \label{fig:alldata}
\end{figure}

Figure \ref{fig:alldata} shows the lifetimes measured by similar
decay curves at a variety of distances from the wire. There are
two series of measurements using two different values of the
spin-flip frequency: $f_0=1.8(1)\,$MHz (circles), and
$f_0=560(10)\,$kHz (squares). The contribution of the transverse
wires to $B_0$ varies slightly with height, but we have measured
this and have compensated for for it by adjusting $B_z$ so as to
maintain constant $f_0$.  In each series, the lifetime for atoms
to remain in the trap exhibits a strong dependence on the distance
from the surface, decreasing by an order of magnitude as the
distance is reduced from $80\,\mu$m to $30\,\mu$m. We note that
the radial trap frequency is by contrast almost constant over this
height range ($735-865\,$Hz for the larger $f_0$ and
$1270-1560\,$Hz for the smaller) because it is inversely
proportional to the distance $r$ from the center of the wire. At a
given height, atoms with a lower spin flip frequency have a
shorter lifetime. The factor of three in $f_0$ produces a factor
of $\sim 5$ in lifetime far from the surface. This ratio decreases
closer to the surface, reaching $\sim 1\frac{1}{2}$ at a distance
of $30\,\mu$m.

The solid (dashed) curve in figure \ref{fig:alldata} shows the
expected thermal spin-flip lifetime $\tau$ for rubidium atoms in
the $|2,2\rangle$ state above a thick plane slab of aluminum
\cite{note1} with $f_0=560\,$kHz ($f_0=1.8\,$MHz). We have
calculated these curves following Henkel {\it et al.}
\cite{henkel99}, whose equation (35) gives the spin flip rate for
a spin-$\frac{1}{2}$ atom. For our case, the rate-limiting
transition is $|2,2\rangle \rightarrow |2,1\rangle$, which has a
smaller matrix element by a factor of $1/\sqrt{5}$. Neglecting an
insignificant height-independent part, $\tau$ is proportional to
$f_0 (h/\delta)(2 h^3+3\delta^3)$, where $h$ is the height above
the surface and $\delta$ is the skin depth of the aluminum at the
frequency $f_0$ of the spin flip transition. The skin depth sets a
natural length scale of $110\,\mu$m ($61\mu$m) for $f_0=560\,$kHz
($f_0=1.8\,$MHz). Close to the surface ($h\ll \delta$), $\tau
\propto h$ and there is no dependence on $f_0$ because $\delta^2
\propto 1/f_0$. By contrast, when $h\gg \delta$, the lifetime is
sensitive to $f_0$ and varies as $h^4 f_0^{3/2}$. There are some
points of striking similarity between our data taken above the
cylindrical wire and this theory for a thick aluminum slab. First,
the lifetimes we observe above the aluminum wire are slightly
longer than those predicted above a plane aluminum slab. This is
what one would expect, since the wire may be viewed as a slab with
some of its more distant parts removed. Second, the lifetime above
the wire increases a little more rapidly than the corresponding
lifetime above a slab. This is a natural consequence of the
additional length scale introduced by the $250\,\mu$m radius of
the wire. Third, the variation with height in the ratio of
lifetimes for the two different values of $f_0$ is echoed in the
predicted behavior above a slab. We interpret these similarities
as confirmation that the loss we observe is indeed due to spin
flips induced by thermal fluctuations of the magnetic field above
the wire.

This result has been achieved after taking considerable care to
control technical noise in the currents being carried by the
wires. In the bandwidth $20\,$Hz - $300\,$kHz, our current
regulation circuits have rms noise below $20\,\mu$A, corresponding
to a part per million of the dc current. At present, however, the
reference voltages that control $B_x$ and the guide wire current
are provided by waveform generators whose outputs do not have this
level of stability.  They increase the total noise current at
audio frequencies by a factor of $\sim 100$. We observe a heating
rate in the range $0.1-0.5\,\mu$K/s that is thought to be the
consequence of this. To look for rf technical noise close to the
spin flip frequency we initially used the antenna that drives the
evaporative cooling as a receiver to search for stray fields. A
spectrum analyzer revealed some very weak rf signals that were
removed after paying attention to ground loops and shielding. The
atoms themselves then proved to be much more sensitive spectrum
analyzers. With the atoms at $225\,\mu$m above the wire, a scan of
$f_0$ revealed a resonant dip in the lifetime at approximately
$1\,$MHz. This was found to be from a parasitic oscillation in the
optical shutter drivers that appeared while the shutters were
closed. Once it was removed, no other rf radiation could be
detected. If there were any residual rf current in the guide wire,
the resulting field would vary as $1/r$, giving a lifetime
proportional to $r^2$, as verified by \cite{leanhardt02a}. This
power law is indicated by the dotted line in figure
\ref{fig:alldata}, arbitrarily placed to pass through one of our
data points. Clearly the height dependence we observe is much
stronger than that. Spin flips induced by rf pickup cause a loss
rate proportional to the rf power at the spin flip frequency
$f_0$. This leads to a constant ratio of lifetimes, not the
height-dependent ratio that we observe.

For the short-range points presented in figure \ref{fig:alldata}
we checked that the lifetime did not change when the size of the
cloud was increased by raising the temperature to $5\,\mu K$. We
are therefore confident that the lifetimes presented here are not
influenced at all by atoms hitting the surface, despite its close
proximity. Presumably the BEC provides reliable spin flip data at
shorter distances than the $27\,\mu$m presented here because it
has a smaller radius than thermal cloud, typically $1-2\,\mu$m.
However, we do not present any shorter range data here because we
have no way other than changing the cloud size to distinguish
surface loss from spin flip loss.

Atom chips aim to control quantum superposition and entanglement
in neutral atoms. From this point of view, it is very undesirable
that thermal fluctuations relax the atomic spins over a few
seconds at a distance of $30\,\mu$m from the surface. This problem
can be reduced if the metallic parts of the chip surface are
restricted to being thin films. Many atom chips are made by
covering a substrate with a $3-5\,\mu$m thick metal layer, which
is then patterned lithographically. Since the fields radiated by
thermal current fluctuations must propagate through the conductor
to reach the atoms, we can consider that the current fluctuations
of interest in a thick slab are localized within a skin depth of
the surface, i.e. in a layer $\sim 100\,\mu$m thick. Hence, metal
layers that are $3-5\,\mu$m thick generate $20-30$ times less
noise power at these frequencies and produces a correspondingly
lower spin flip loss rate. Recently we have used a different
apparatus to hold rubidium atoms in a magnetic trap $30\,\mu$m
away from a $200\,$nm-thick gold film. We find that the thin gold
film does indeed produce a much lower spin flip rate
\cite{retter02} than we see here at a distance of $30\,\mu$m from
the wire. Leanhardt {\it et al.}, \cite{leanhardt02a} have also
observed long lifetimes at $70\,\mu$m from a $5\,\mu$m copper
film.

Even so, the coupling between the atoms and the substrate poses an
important technical difficulty for atom chips if the atoms are to
approach the surface much more closely than $\sim 10\,\mu$m. One
method could be to cool the chip to suppress the thermal
fluctuations, and to use superconducting wire to avoid dissipating
heat. An alternative is to avoid current-carrying wires altogether
on the surface of the chip by using permanent magnets to produce
the required microscopic structures. Videotape provides a simple
way of making structures down to $\sim5\,\mu$m in size
\cite{hindsreview99}. In our laboratory we have loaded atoms into
microtraps formed above sinusoidally magnetized videotape, where
they can readily be cooled by evaporation and manipulated by
externally applied magnetic fields \cite{retter02,retter03}. In
order to reach an even smaller length scale, we are now exploring
the use of magneto-optical films for making magnetic traps. These
are metallic, but can be as thin as $30\,$nm and still produce
traps several mK deep a few $\mu$m away from the surface.

We are indebted to Alan Butler for expert technical assistance. We
also acknowledge valuable discussions with Gabriel Barton and
Jozsef Fortagh. This work was supported by the EPSRC and PPARC
research councils of the UK and by the ACQUIRE and FASTNET
networks of the European Union.


\begin{thebibliography}{12}
\bibitem{weinstein95}J. D. Weinstein and K. G. Libbrecht, Phys. Rev. A {\bf52}, 4004, (1995).
\bibitem{vuletic98}V. Vuletic, T. Fischer, M. Praeger, T. W.
H\"{a}nsch and C. Zimmermann, Phys. Rev. Lett. {\bf80}, 1634,
(1998).
\bibitem{fortagh98}J. Fortagh, A. Grossmann, C. Zimmerman and T. W. H\"{a}nsch, Phys. Rev. Lett. {\bf81}, 5310, (1998).
\bibitem{mueller99}D. M\"{u}ller and D. Z. Anderson and R.J. Grow and P. D. D. Schwindt and E. A. Cornell, Phys. Rev. Lett. {\bf83}, 5194, (1999).
\bibitem{dekker00}N. H. Dekker {\it et al.}, Phys. Rev. Lett. {\bf84}, 1124, (2000).
\bibitem{key00}M. Key {\it et al.}, Phys. Rev. Lett. {\bf87}, 230401, (2000).
\bibitem{hindsreview99}E. A. Hinds and I. G. Hughes, J. Phys. D. {\bf32}, R119, (1999).
\bibitem{folmanreview02}R. Folman and P. Kr\"{u}ger and C. Henkel and J. Schmiedmayer, Adv. Atom. Mol. Opt. Phys {\bf48}, 263, (2002).
\bibitem{hinds01}E. A. Hinds and C. J. Vale and M. G. Boshier, Phys. Rev. Lett. {\bf86}, 1426, (2001).
\bibitem{haensel01}W. H\"{a}nsel and J. Reichel and P. Hommelhoff and T. W. H\"{a}nsch, Phys. Rev. A {\bf64}, 063607, (2001).
\bibitem{andersson02}E. Andersson {\it et al.}, Phys. Rev. Lett. {\bf88}, 100401, (2002).
\bibitem{calarco00}T. Calarco {\it et al.}, Phys. Rev. A. {\bf61}, 022304, (2000).
\bibitem{horak02}P. Horak {\it et al.}, quant-ph/0210090, (2002).

\bibitem{fortagh02}J. Fortagh and H. Ott and S. Kraft and A. Gunther and C. Zimmerman, Phys. Rev. A {\bf66}, 041604(R), (2002).
\bibitem{leanhardt02}A. E. Leanhardt {\it et al.}, Phys. Rev. Lett. {\bf89}, 040401, (2002).
\bibitem{kraft02}S. Kraft {\it et al.}, J. Phys. B {\bf35}, 469, (2002).

\bibitem{leanhardt02a}A. E. Leanhardt {\it et al.}, cond-mat/0211345, (2002).
\bibitem{henkel99}C. Henkel and S. P\"{o}tting and M. Wilkens", Appl. Phys. B {\bf69}, 379, (1999).
\bibitem{jones02}M. Jones, D.Phil Thesis, Sussex University,(2002).
\bibitem{note1}The corresponding curves for copper are almost the same because the two metals have very similar resistivities: $2.6 \times 10^{-8}\,\Omega\,$m for Al and $1.7 \times 10^{-8}\,\Omega\,$m for Cu.
\bibitem{retter02}J. Retter, D.Phil Thesis, Sussex University, (2002).
\bibitem{retter03}J. Retter {\it et al.}, to be published, (2003).
\end{thebibliography}
\end{document}